\documentclass[aps,prd,superscriptaddress,12pt,notitlepage]{revtex4}
\usepackage{graphics}
    \usepackage{graphicx}
    \usepackage{amsmath,amssymb,mathrsfs}
\usepackage{epsf}
\usepackage{epsfig}
\usepackage[font=small,skip=0pt,justification=raggedright]{caption}
\usepackage{xcolor}

\newcounter{fig}   \newcommand{\lbfig}[1]{\refstepcounter{fig}
\label{#1} }

\newcommand{\Tr}{{\rm Tr}}

\newcommand{\bea}{\begin{eqnarray}}
\newcommand{\eea}{\end{eqnarray}}
\newcommand{\be}{\begin{equation}}
\newcommand{\ee}{\end{equation}}

\newcommand{\re}[1]{(\ref{#1})}

\newcommand{\pa}{\partial}

\newcommand{\eqn}{\begin{eqnarray}}
\newcommand{\eqnx}{\end{eqnarray}}

\begin{document}

\title{$SU(2)$ Yang-Mills solitons  in $R^2$ gravity}

\author{I. Perapechka}
\affiliation{Department of Theoretical Physics and Astrophysics, BSU, Minsk 220004, Belarus}
\author{Ya. Shnir}
\affiliation{Department of Theoretical Physics and Astrophysics, BSU, Minsk 220004, Belarus\\
BLTP, JINR, Dubna 141980, Moscow Region, Russia\\
Department of Theoretical Physics, Tomsk State Pedagogical University, Russia}

\begin{abstract}
We construct new family of  spherically symmetric regular solutions of $SU(2)$ Yang-Mills theory coupled to
pure $R^2$ gravity. The particle-like field configurations possess non-integer non-Abelian magnetic charge.
A discussion of the main properties of the solutions and their differences from the usual
Bartnik-McKinnon solitons in the asymptotically flat case is presented. It is shown that there is
continuous family of linearly stable non-trivial solutions in which the gauge field has no nodes.
\end{abstract}
\maketitle

\section{Introduction}

Modified theories of gravity gained increasing interest in past decade since
it now seems accepted that the inflationary scenario in the early Universe is related with
modification of the usual Einstein-Hilbert action  \cite{Starobinsky:1980te}, in particular via
addition of the quadratic curvature terms. The simplest $R+R^2$ model \cite{Salam:1978fd} is proven to be
renormalizable \cite{Stelle:1976gc,Stelle:1977ry}, further, it appears in a natural way as a limit of the
string theory \cite{Ellis:2013xoa}. Such generalizations have been also studied as an explanation for the
dark energy problem \cite{Capozziello:2002rd,Starobinsky:2007hu}.

The linear Einstein term is not always assumed to be present in the action of the modified gravity. The pure
$R^2$ theory has some advantages \cite{Kounnas:2014gda,Alvarez-Gaume:2015rwa}, in particular it is the
only ghost-free higher order theory.
On the other hand, it admits supergravity generalization \cite{Ferrara:2015ela,Kuzenko:2016nbu}.
Further, pure $R^2$ black hole and wormhole solutions were constructed in
\cite{Kehagias:2015ata,Duplessis:2015xva}, very recently the
pure $R^2$ theory supplemented by a set of complex scalar fields was investigated in \cite{Ellis:2017xwz} as
a limit of supergravity model. However, not much known about solutions of the $R^2$ gravity
coupled to the non-Abelian fields.

Spatially localised particle-like solutions of the classical Yang-Mills theory coupled to the usual gravity
have been the subject of long standing research interest
since Bartnik and McKinnon found these solutions in 1988 \cite{Bartnik:1988am}.
These globally regular self-gravitating field configurations were discovered numerically in the asymptotically
flat $\mbox{SU}(2)$ Einstein-Yang-Mills (EYM) theory. It has been shown that they are linked to the
nontrivial hairy black holes \cite{Volkov:1989fi,Bizon:1990sr}, this observation sparked a lot of activity over last two
decades, see e.g. \cite{Volkov:1998cc,Volkov:2016ehx}.
The $\mbox{SU}(2)$ Bartnik-McKinnon (BM) solutions in the asymptotically flat space are spherically symmetric and
purely magnetic with the net magnetic charge equal to zero
\cite{Galtsov:1989ip,Bizon:1992pi}, further they are unstable with respect to linear
perturbations of the metric and the gauge field \cite{Straumann:1989tf,Zhou:1991nu,Lavrelashvili:1994rp}.
The
BM solutions were subsequently generalized to the $\mbox{SU}(N)$ \cite{Kuenzle:1991wa,Brodbeck:1994np,Kleihaus:1995tk} and the $\mbox{SO}(N)$
\cite{Bartnik:2009hc,Oliynyk:2000vf} Einstein-Yang-Mills theory,
axially symmetric generalizations of the BM solutions were considered in
\cite{Kleihaus:1996vi,Ibadov:2004rt}.

An interesting observation is that
a variety of features of asymptotically flat self-gravitating BM solutions
and the corresponding hairy black holes are not shared by
their counterparts in the asymptotically anti-de Sitter (AdS) space-time
\cite{Winstanley:1998sn,Bjoraker:1999yd,Bjoraker:2000qd,Sarbach:2001mc}.
There is a continuum of new magnetically charged field configurations
with asymptotically non-vanishing magnetic flux, which are stable under linear perturbations of the fields.
One can consider these solutions as describing
non-Abelian monopoles in the absence of a Higgs field with a non-integer magnetic charge
\cite{Bjoraker:1999yd,Bjoraker:2000qd}.
On the other hand, these solutions are  relevant in the context of the AdS/QFT holographical  correspondence
\cite{Gubser:2008wv}.
As discussed in \cite{Baxter:2007au,Winstanley:2008ac}, the EYM solutions in AdS$_4$ spacetime possess
generalizations with higher gauge groups, there is also variety of interesting
axially-symmetric AdS solutions of the EYM equations \cite{Radu:2001ij,Kichakova:2014fta}.

The main purpose of this work is to explicitly construct $R^2$ counterparts of the spherically symmetric
solutions of the EYM system, looking for new features induced by the different structure of the gravitational
part of the action.

\section{$R^2$ Yang-Mills model}
We consider the $\mbox{SU}(2)$ Yang-Mills gauge field coupled to pure $R^2$ gravity in $(3+1)$ dimensions.
The model is defined by the scale invariant action
\be
\label{act}
S=\int \sqrt{-g}\left(\frac{R^2}{2\kappa} - \Tr\; F_{\mu \nu}F^{\mu \nu} \right)d^4x
\ee
where $R$ is the usual curvature scalar,  $g$ denotes the determinant
of the metric $g_{\mu \nu}$ and $\kappa$ is the effective gravitational coupling
constant. The matter field sector is defined by the $\mbox{SU}(2)$ field strength tensor
$$
F_{\mu \nu} = \pa_\mu A_\nu - \pa_\nu A_\mu - i [A_\mu,A_\nu] \, ,
$$
and $A_\mu = \frac12 A_\mu^a \tau^a \in  \mathfrak{su}(2)$.

It is known the pure $R^2$ theory is equivalent to the usual Einstein gravity with additional real scalar field
\cite{Kounnas:2014gda}. Indeed,  one can replace the  $R^2$ term
with $\frac{R^2}{\kappa} \to 2tR - \kappa t^2 $, where $t$ is a Lagrange multiplier.  Then the variational equation
for the field $t$, which is non-propagating in this frame, yields the $R^2$ term back. Consequent rescaling of the
metric to the Einstein frame and redefinition of the field $t \to \phi \sim \ln (2t) $, transforms
the pure $R^2$ gravity to the standard gravitational action with a cosmological constant,
coupled to a massless scalar field $\phi$ \cite{Kounnas:2014gda}.

The scale invariant model \re{act} in the Einstein frame  after rescaling takes the form
\be
S=\int \sqrt{-g}\left(\frac12 R - \frac12\partial_\mu \phi\partial^\mu \phi - \frac{\kappa}{8} -
\Tr\; F_{\mu \nu}F^{\mu \nu} \right)d^4x
\label{actJord}
\ee
Here the quantity $\frac{\kappa}{4}$ is playing the role of the cosmological constant.

Such a theory, with positive cosmological constant and both scalar and non-Abelian Yang-Mills
fields in the matter sector, is not very common,
Most attention is usually devoted to similar models with an exponential dilaton coupling, see e.g.
\cite{Lavrelashvili:1992ia,Kleihaus:1996vi,Radu:2004xp}. On the other hand, the reformulated
model in the Einstein frame may only capture part of the possible solutions of the original theory with
$R^2$ term \cite{Kehagias:2015ata},
so hereafter we restrict our consideration to the model \re{act}.

Variation of the action \re{act} with respect to the metric $g_{\mu \nu}$  yields the $R^2$ gravity equations,
which are counterparts of the usual Einstein equations:
\be
R R_{\mu\nu} - \frac14 R^2g_{\mu \nu} - \nabla_\mu \nabla_\nu R + g_{\mu \nu} \square R = \frac{\kappa}{2}
T_{\mu \nu} \, .
\label{eqs-R2}
\ee
Here the Yang-Mils stress-energy tensor is
\be
T_{\mu \nu} = -4\; \Tr \left(F_{\mu \rho} F_{\nu}^{\rho} -\frac14 g_{\mu \nu} F^{\rho \sigma}
F_{\rho \sigma} \right) \, .
\label{stress-energy}
\ee
Variation of the action \re{act} with respect to the gauge field $A_\mu$
leads to the Yang-Mills equations in the curved space-time
\be
\nabla_\mu F^{\mu\nu} - i [A_\mu, F^{\mu\nu}]=0
\label{eqs-YM}
\ee
Note that the equations of $R^2$ gravity \re{eqs-R2} are highly
non-linear fourth order differential equations, it is not obvious
how such a system can be integrated in a general case.
However, we can see that the left hand side of the gravitational equations \re{eqs-R2}
is covariantly constant, thus \cite{Kehagias:2015ata}
\be
\nabla^\mu T_{\mu \nu} = 0 \, .
\ee
Further, the action \re{act} is classically scale invariant, i.e. $\Tr\; T_{\mu \nu} = 0$. Taking trace
on both sides of the Eq.~\re{eqs-R2} we obtain
\be
\square R = 0 \, .
\label{boxR}
\ee
Thus, in the static case the regular solutions of the Laplace equation \re{boxR} on the entire space
$\mathbb{R}^3$ without event horizon,
are harmonic functions. In such a case the Liouville's theorem guarantees that
$R=\textrm{const}$ is a solutions of Eq.~\re{boxR}.
This result greatly simplifies the consideration, indeed
there are two distinct situations. In the case of zero curvature,
the solutions of the model \re{act} in the asymptotically
flat spacetime are trivial and $T_{\mu \nu} = 0$.
In the second case we suppose that the scalar curvature is a non-vanishing constant.
Then
the $R^2$ gravitational equation \re{eqs-R2} can be written in the
Einsteinian form
\be
R_{\mu\nu} - \frac14 g_{\mu\nu} R = \frac{\kappa}{2 R} T_{\mu\nu} \, .
\label{gr-red}
\ee
Since the curvature scalar is a constant, solutions of this equation are given by the
equivalent Einstein equations in the AdS spacetime upon identification $R=4\Lambda$ and rescaling of the
gravitational coupling constant as $\kappa \to \frac{\kappa}{8\Lambda}$.

\subsection{Spherically symmetric ansatz and the boundary conditions}
We restrict our consideration to the static spherically symmetric field configuration, which are counterparts of the usual
BM solutions. Then the spherically symmetric purely magnetic
Ansatz for the Yang-Mill field is given by
\be
A_0 =0\, ; \qquad A_i^a = \varepsilon^a_{ij}\frac{x_j}{r^2}\left(1-w(r) \right) \, ,
\label{YM}
\ee
where $w(r)$ is the profile function. For the metric, we employ the usual
spherically symmetric  line element
\be
ds^2=-\sigma^2(r)N(r)dt^2 +\frac{dr^2}{N(r)} + r^2\left(
d\theta^2 +\sin^2 \theta ~ d\phi^2
\right) \, .
\label{metric}
\ee

Within this specific ansatz \re{metric}, the Laplace equation \re{boxR} becomes
\be
\frac{R_{rr}}{R_r}+\frac{N_{r}}{N}+\frac{\sigma_{r}}{\sigma} + \frac{2}{r} = 0 \, .
\ee
A general solution of this equation can be written as
$$
R=C_1 + C_2 \int\limits_0^\infty \frac{dr}{r^2 \sigma N} \, ,
$$
where $C_1,C_2$ are two arbitrary constants. The regularity condition yields $C_2=0$,
as it is mentioned above, the curvature scalar is a constant.

Note that in the case of positive constant curvature, the spherically-symmetric
solutions of the Yang-Mills system coupled to $R^2$ gravity are effectively
equivalent to the asymptotically de-Sitter solutions in the Yang-Mills model coupled to the usual
Einstein gravity \cite{Volkov:1996qj}.
This is not the case, however, for the asymptotically AdS solutions
with a negative constant curvature, $R<0$. In the Einstein gravity this situation would
correspond to a non-conventional choice of the negative gravitational coupling constant.
Hereafter we consider the case of $R<0$.

As we can see, this equivalence also holds for a theory \re{actJord} in the Jordan frame.
Indeed, in the  case of constant scalar curvature the scalar field $\phi$ should also be constant, the
corresponding dynamical equation is just $\square \phi=0$ and the field $\phi$ is
constant everywhere in space.  Thus, it does not affect the dynamical equations for the metric functions and
the  Yang-Mills field, in the Jordan frame the scalar field is effectively decoupled.

Within the spherically symmetric ansatz \re{YM},\re{metric}, the
variational equations associated with the action \re{act} can be
reduced to the following system of three non-linear differential
equations
\be
\begin{split}
w_{rr}&=w_r \frac{r^2R[r^2 R +4(N-1)] +2\kappa (w^2-1)^2}{4NR r^3}
+\frac{w(w^2-1)}{Nr^2}\, , \\
N_r&= -\frac{\kappa (w^2-1)^2}{2Rr^3}+\frac{R[Rr^2 +4(N-1)]+4\kappa N w_r^2}
{4Rr} \, , \\
\sigma_r&=\frac{\kappa \sigma w_r^2}{Rr} \, .
\end{split}
\label{eqs-sys}
\ee
Note that the metric function $\sigma(r)$ can be integrated out,
$$
\sigma(r) = \sigma(0) \exp\left(\frac{\kappa}{R} \int_0^\infty \frac{w_r^2}{r} dr \right) \, .
$$

The series expansion of the equations \re{eqs-sys} near the origin yields
\be
\begin{split}
w&\approx 1 - br^2 + O(r^4)\, , \\
\sigma& \approx \sigma_0 + \frac{2\kappa b^2}{R}+ O(r^4)\, ,\\
N& \approx 1- \left(\frac{2\kappa b^2}{R} + \frac{R^2}{12} \right) r^2 + O(r^4) \, .
\end{split}
\ee
Similarly, on the spacial boundary
\be
\begin{split}
w&\approx  w_\infty - \frac{6w_\infty (w_\infty^2-1)}{Rr^2} +  O(r^{-4})\, ,\\
\sigma& \approx 1 +  O(r^{-4})\, ,\\
N& \approx 1 -\frac{R}{12}r^2 - \frac{2M}{r} +\frac{\kappa(w_\infty^2-1)^2}{2Rr^2}
+ O(r^{-3}) \, .
\end{split}
\ee
Here $\sigma_0,M$ and $w_\infty$ are constants that
have to be determined numerically. The constants $\kappa$ and $b$ are the parameters of
a particular spherically-symmetric solutions of the $R^2$ Yang-Mills coupled system.

Thus, in order to obtain regular solutions of this model with finite energy density
we have to impose the following boundary conditions
\be
w(0) = 1\, ,\quad w_r(0) = 0\, ,\quad N(0)=0\, \quad \sigma(\infty)=1 \, .
\label{bc}
\ee
Evidently, they agree with the corresponding boundary conditions for the asymptotically AdS
EYM system, see \cite{Bjoraker:1999yd}.

The static regular localized solutions  are characterized by the mass $\mathcal{M}$ and
by the non-Abelian magnetic charge
\be
Q=\frac{1}{4\pi} \int \sqrt{-g}\varepsilon^{ijk}F_{ij} dS_k  = (1-w_\infty^2)\frac{\tau_3}{2} \, .
\label{Q}
\ee
For the gauge invariant charge we use the definition $|Q|$, where the vertical bars denote the Lie-algebra norm
\cite{Sudarsky:1992ty}.
Similar to the asymptotically AdS solutions in the usual EYM system
\cite{Winstanley:1998sn,Bjoraker:1999yd,Bjoraker:2000qd,Sarbach:2001mc},
the function $w(r)$ does not need to have an asymptotic value $w_\infty = \pm 1$, the charge
\re{Q} generally is not integer.

Note that by analogy with the corresponding solutions in the AdS$_4$ spacetime,  we can
reparametrize the metric function $N(r)$ as
\be
N(r)=1-\frac{2m(r)}{r} -\frac{R r^2}{12} \, ,
\label{metricN}
\ee
where the function $m(r)$ has an asymptotic limit $m(\infty) = M$.
However, in the $R^2$ gravity the usual definition of the Arnowitt-Deser-Misner mass
should be modified in the comparison with the case of the conventional general relativity \cite{Deser:2002rt}.
Thus, the parameter $M$ does not uniquely specify the mass, in the system under consideration
the static energy is defined as $\mathcal{M} = MR$ instead \cite{Deser:2002rt,Deser:2007vs}.

\section{Numerical results}

Solutions for system of equations \re{eqs-sys}  with boundary conditions \re{bc} are constructed
numerically using shooting algorithm, based on Dormand-Prince 8th order method with
adaptive stepsize. The relative errors of calculations are lower than $10^{-10}$.
Similar to the case of the soliton solutions in the usual EYM model, for each fixed values of
the parameter $\kappa$  there is a continuous set of regular magnetic solutions labeled by the
free adjustable parameter $b$. Typical solutions are displayed in Fig.~\ref{f-1}.
For all the solutions we present we make use of the scale invariance
of the model \re{act} and take the value of the curvature scalar $R=-1$.

\begin{figure}[hbt]
\lbfig{f-1}
\begin{center}
\includegraphics[height=.28\textheight, angle =0,trim = 60 20 80 50, clip = true]{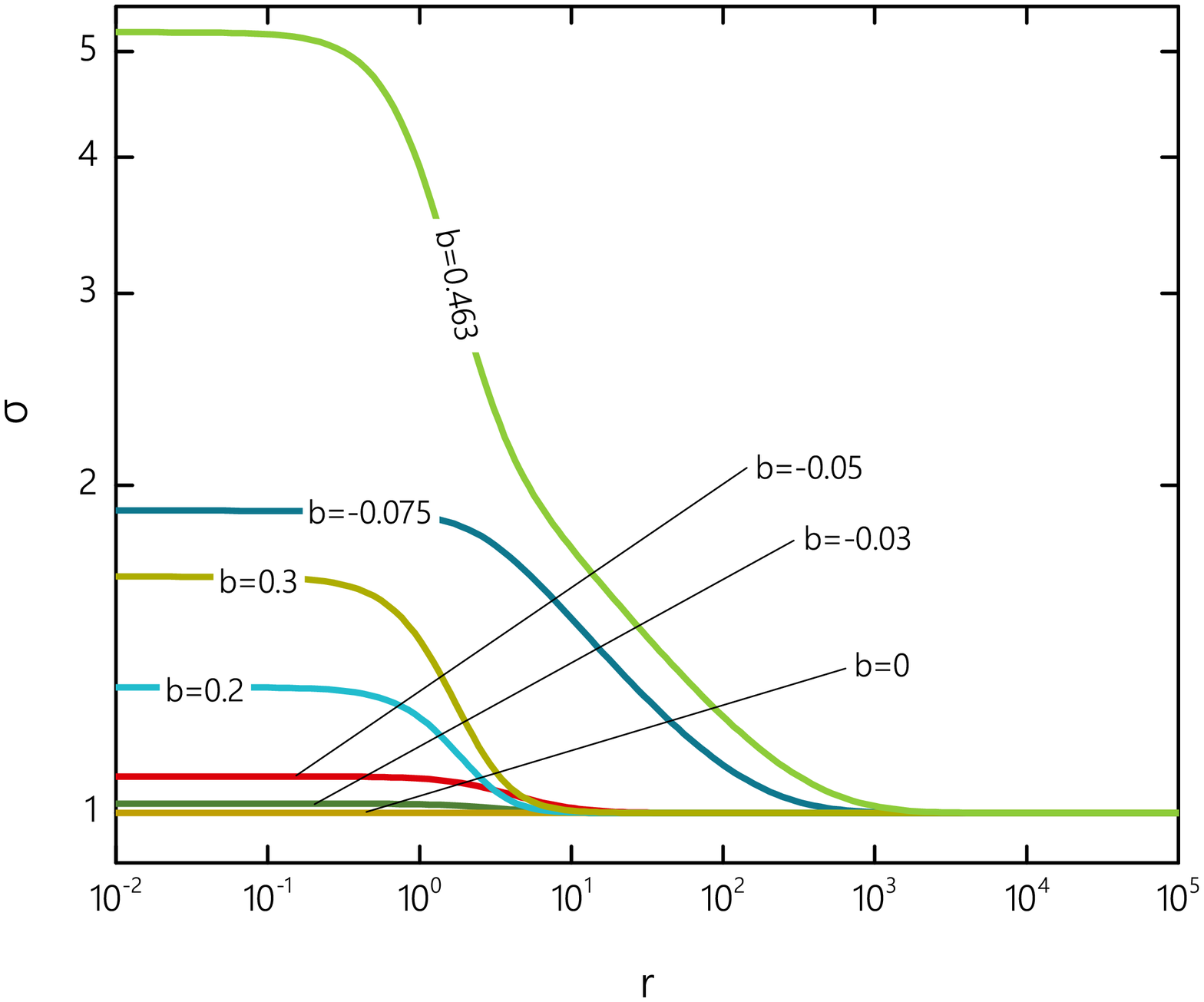}
\includegraphics[height=.28\textheight, angle =0,trim = 60 20 80 50, clip = true]{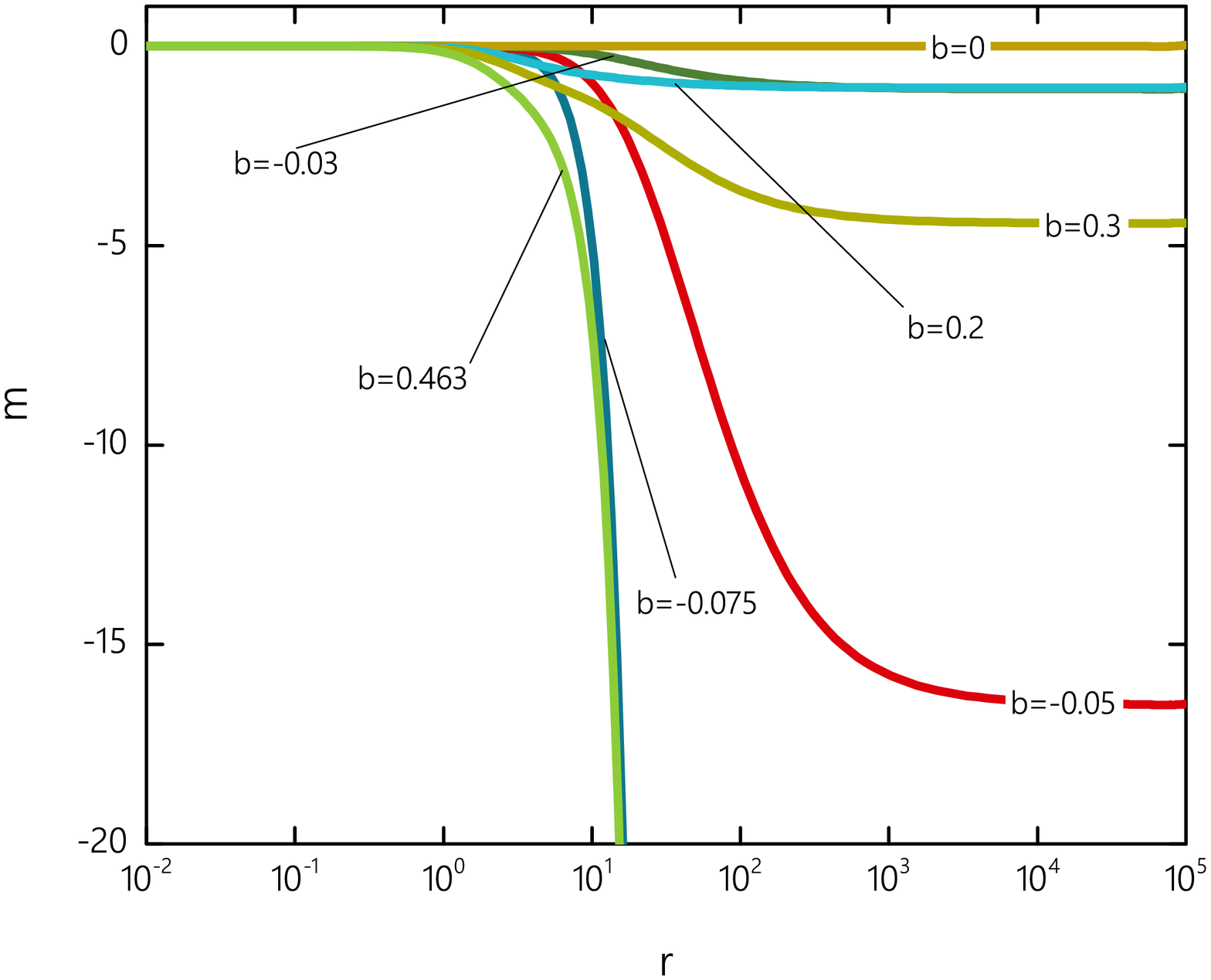}
\includegraphics[height=.28\textheight, angle =0,trim = 60 20 80 50, clip = true]{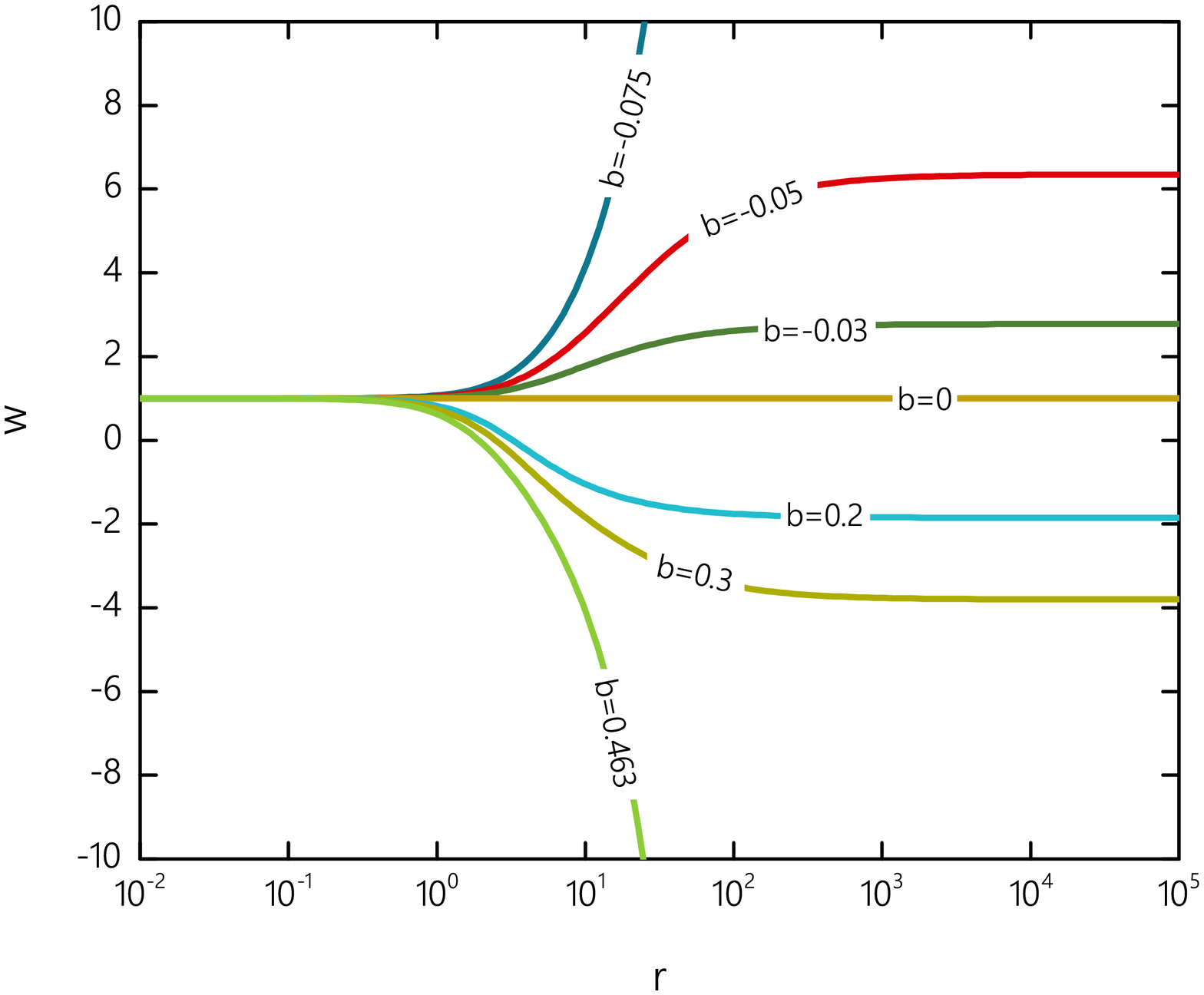}
\end{center}
\caption{\small
$R^2$ EYM solutions for $R=-1$, $\kappa=1$ and some set of values of $b$:
The  metric function $\sigma(r)$ (upper left plot), the mass function $m(r)$ (upper right plot) and
the profile function of the gauge field $w(r)$ (bottom plot) are plotted as functions of the radial coordinate $r$.
}
\end{figure}

Variation of the shooting parameter $b$ gives us a continuous family of solutions, which
are qualitatively similar to the usual Einstein-Yang-Mills monopoles in
asymptotically AdS spacetime \cite{Bjoraker:1999yd,Bjoraker:2000qd}.
However, there are some important differences. First, the regular finite energy $R^2$ EYM magnetic
solutions exist only for one finite interval of values of the parameter $b$ bounded from above and below,
$b\in [b_{min}, b_{max}]$ where $b_{min}<0$.  Contrary to this case,
a continuum of monopole solutions in the conventional EYM AdS$_4$ exists for all values of the shooting
parameter bounded from above only \cite{Bjoraker:1999yd,Bjoraker:2000qd}. On the other hand,
the family of finite energy EYM solutions in the fixed AdS background also exist
for only one interval in parameter space \cite{Radu:2001ij}.
Secondly, there is an additional parameter labeling the usual EYM AdS$_4$ solutions,
the number $n$ of oscillations of the Yang-Mills field. Our numerical results show that
for the $R^2$ EYM system there are only solutions with $n=0$ or $n=1$.
Thus, although the $R^2$ gravity coupled to the Yang-Mills theory can be
conformally transformed to Einstein frame, where it
takes the form of the standard Einstein gravity with cosmological constant and
both massless scalar and non-Abelian Yang-Mills fields in the matter sector, the solutions of these models
are still quite different, there is no one-to-one correspondence between them.

Setting the shooting parameter $b=0$ yields the trivial zero energy solution  with vanishing non-Abelian magnetic field
in the AdS space with a cosmological constant $\Lambda = \frac{R}{4}$.
Increasing of the parameter $b$ lead to increase of both the energy and the
magnetic charge of the configuration.
These  solutions are of particular interest because, as we will see below, they are stable against linear perturbations.
At some value of $b$ the asymptotic value of the gauge field function $w_\infty$
approaches zero and the magnetic charge takes its maximal value, $|Q|=1$.  Further increase of the parameter $b$ leads to
decrease of the charge which again deviates from an integer,
on this unstable branch of solutions the gauge field profile function $w(r)$  has a single node.
At some upper critical value of the parameter $b$ both the mass and the energy diverge, see Fig.~\ref{f-2}

\begin{figure}[hbt]
\lbfig{f-2}
\begin{center}
\includegraphics[height=.28\textheight, angle =0,trim = 60 20 80 50, clip = true]{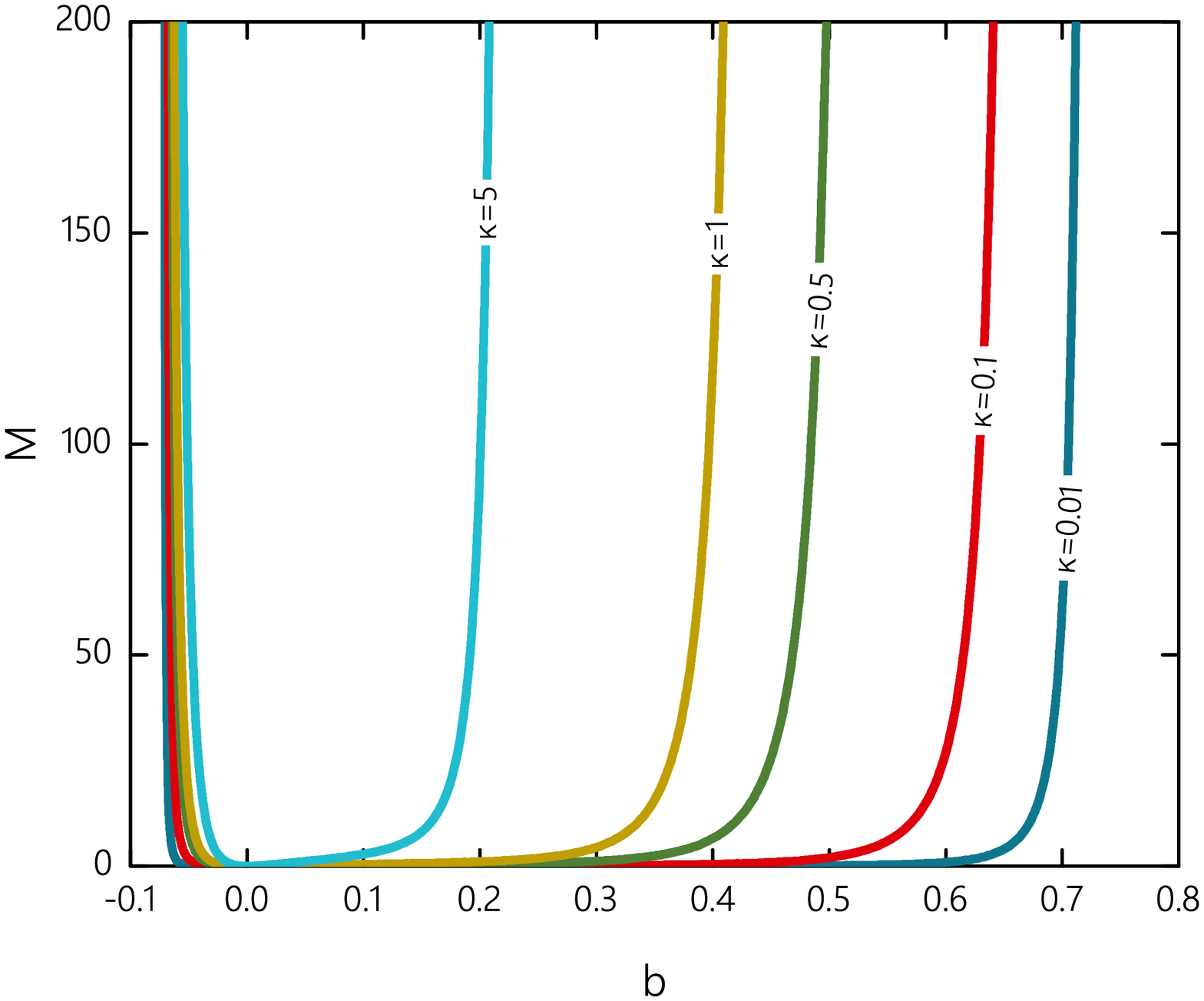}
\includegraphics[height=.28\textheight, angle =0,trim = 60 20 80 50, clip = true]{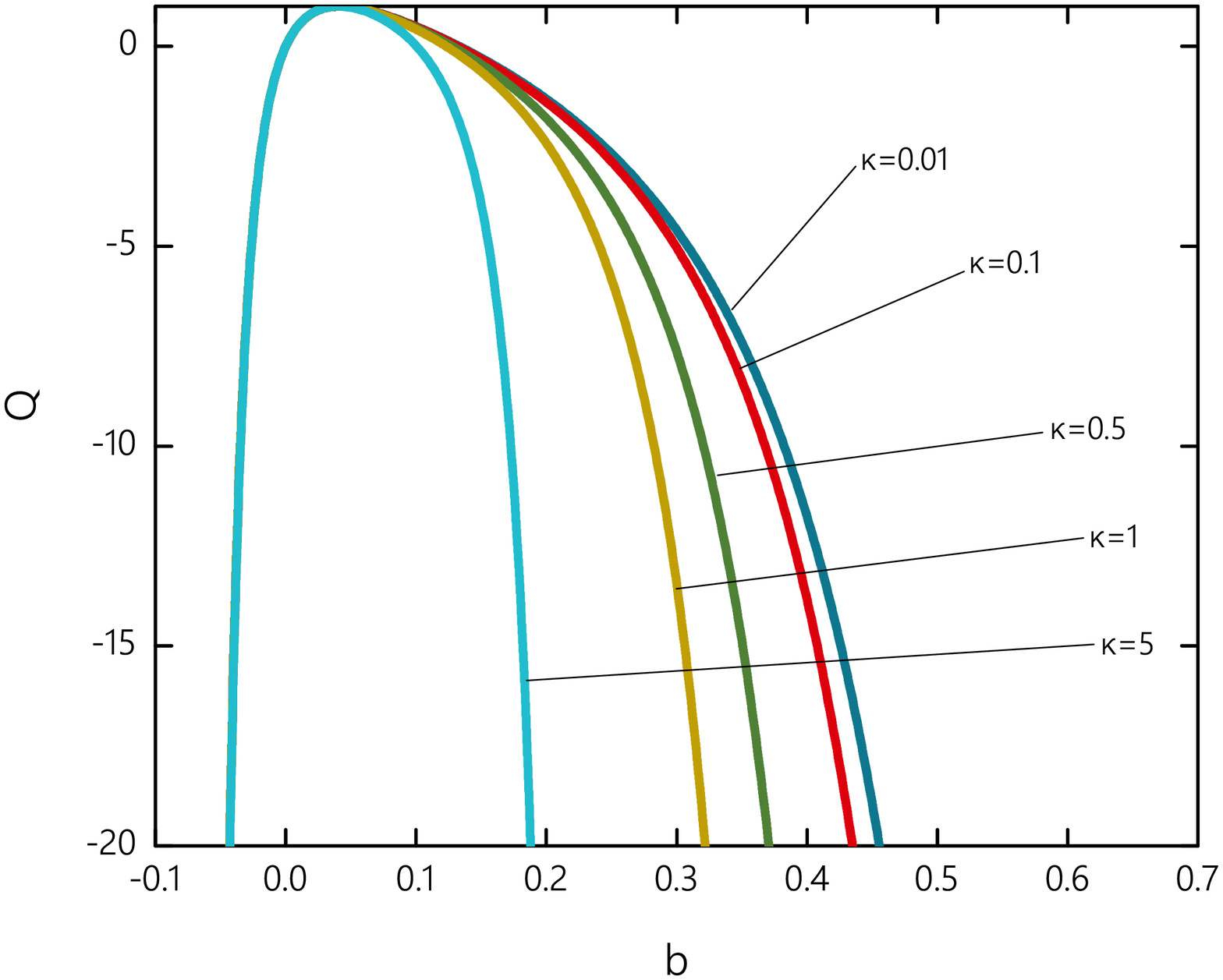}
\includegraphics[height=.28\textheight, angle =0,trim = 60 20 80 50, clip = true]{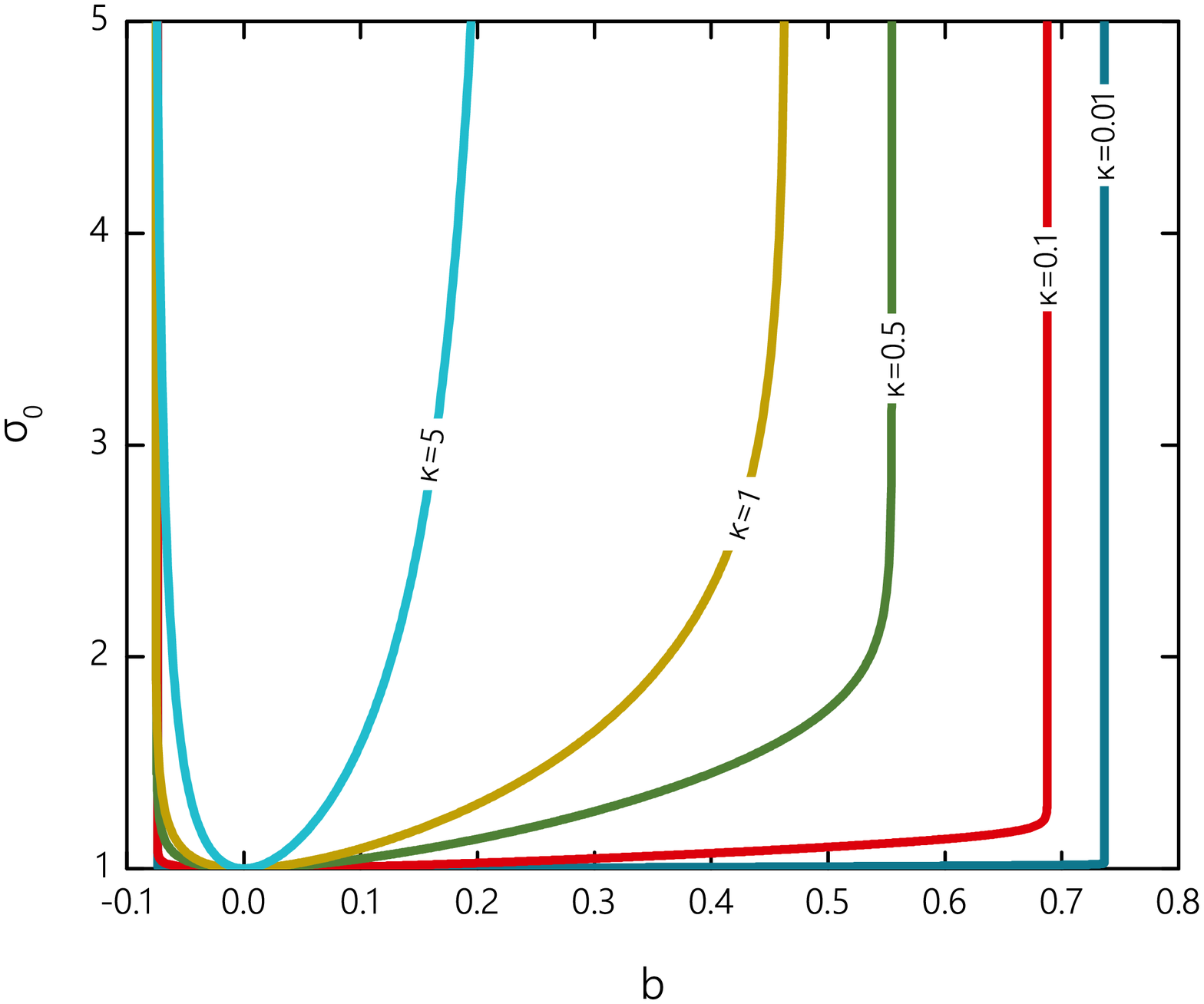}
\end{center}
\caption{\small The energy (upper left plot), the magnetic charge (upper right plot) and the value of the
metric function $\sigma_0=\sigma(0)$ (bottom plot) of $R^2$ EYM solutions are shown as functions
of the parameter $b$ for some set
of values of the gravitational coupling $\kappa$ at $R=-1$.
}
\end{figure}

Similarly, decreasing of the parameter $b$ from zero is leading to decrease of the magnetic charge of the configuration,
along this branch the energy rapidly increases, both the energy and the charge diverge at some negative value of
$b=b_{min}$. No solution seems to exist for $b$ less than this  value. Note that the pattern of critical behavior is different
from the usual EYM monopoles, the metric function $N$ of the $R^2$ EYM solutions diverge at both extremities of the interval
of values of $b$, while it approaches zero in the case of the EYM AdS$_4$ solutions \cite{Bjoraker:2000qd}.
As shown in Fig.~\ref{f-2}, the interval of values of the parameter $b$
decreases when the gravitational coupling constant $\kappa$ gets larger. The contraction of this interval is mainly
because of decrease of the upper critical value $b_{max}$, the lower critical value $b_{min}$
weakly depends on variations of the coupling constant $\kappa$. Dependencies of the critical values of the
parameter $b$ on $\kappa$ are shown in Fig.~\ref{bcr}.
Both critical values, $b_{max}$ and $b_{min}$, approach zero
as $\kappa$ tends to infinity, however the upper critical value $b_{max}$ decreases monotonically,
while the lower critical value $b_{min}$ possesses a minimum at $\kappa\sim 6$.

\begin{figure}[hbt]
\lbfig{bcr}
\begin{center}
\includegraphics[width=.6\textwidth, angle =0,trim = 60 20 80 50, clip = true]{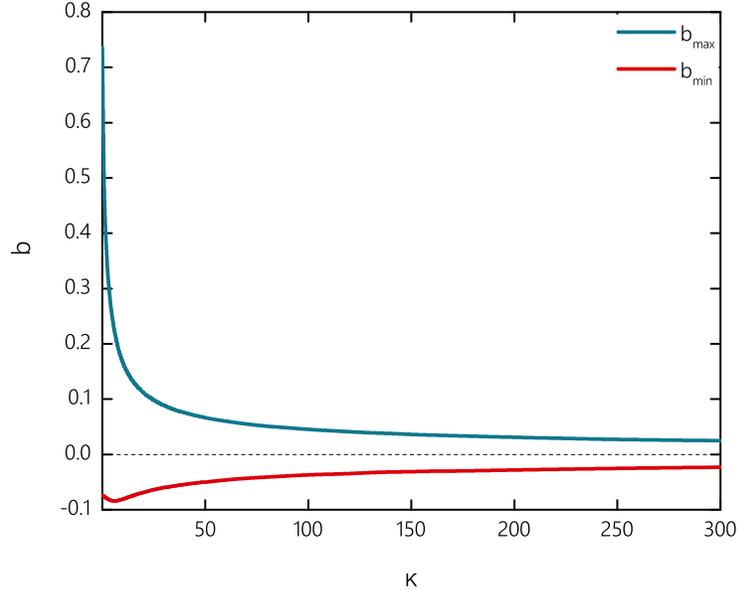}
\end{center}
\caption{\small Critical values of the parameter $b$ are shown as functions of
the effective gravitational coupling constant $\kappa$ for $R=-1$.
}
\end{figure}

\section{Linear stability analysis}
In is known that the usual EYM AdS$_4$ magnetic solutions with
no node in  $w(r)$ are stable with respect to linear perturbations \cite{Bjoraker:1999yd,Bjoraker:2000qd}, so
we can expect the same arguments can be applied to the corresponding solutions of the $R^2$ EYM system.
We consider small time-dependent perturbations of the configuration \re{YM} described by the general magnetic
ansatz for the spherically symmetric Yang-Mills connection
\be
A = \frac12 \left\{u(r,t) \tau_3 dr + [w(r,t) \tau_1 + v(r,t)\tau_2] d\theta + [w(r,t)\tau_2
- v(r,t) \tau_1 +\cot\theta \tau_3]\sin\theta
d\phi \right\} \, .
\ee
It is more convenient in the stability analysis to make use of
the following parametrization for the metric
\be
ds^2 = -e^{\nu(r,t)} N(r,t)dt^2 + e^{\lambda(r,t)} dr^2 + r^2(d\theta^2 +\sin^2\theta d\phi^2) \, ,
\ee
instead of \re{metric}.

The functions in this general ansatz can be now written as the sum of the static solution,
which stability we are investigating, and a time dependent perturbations:
\be
\begin{split}
w(r,t) &= w(r)+ \delta w(r,t)\, , \qquad
u(r,t) =\delta u(r,t)\, , \qquad
v(r,t) =\delta v(r,t)\, ; \\
\nu(r,t) &= \nu(r)+ \delta \nu(r,t)\, , \qquad
\lambda(r,t) = \lambda(r)+ \delta \lambda(r,t)\, .
\label{pert}
\end{split}
\ee
Substituting \re{pert} into the action of the system \re{act} and
retaining only terms linear in perturbations, we variate the corresponding functional with respect to
the fluctuations of the matter fields $\delta w, \delta u$ and $\delta v$. The equations for
fluctuations of the metric fields can be obtained from the linearized gravitational equations \re{gr-red}.
In particular, integration of the corresponding $rt$-equation over time yields the relation
$$
\delta \lambda = \frac{2\kappa w_r}{Rr} \delta w \, .
$$
Another relation between the perturbations
can be obtained from the linear combination of the $tt$- and
$rr$- Einsteinian equations, together with the corresponding unperturbed equations:
$$
\delta\lambda_r + \delta\nu_r = \frac{4\kappa w_r}{Rr} \delta w \, .
$$
With there relations at hands, we arrive to the following system of the linearized equations
\be
\begin{split}
&4r^4R^2 e^\lambda  \delta w_{tt} +
2\kappa e^{\lambda+\nu} \delta w w_r^2 \left( r^2 R(r^2R-4) +2\kappa(w^2-1)^2\right)\\
&+4r^3R^2e^\nu(\delta w_r - r \delta w_{rr})+
rR e^{\lambda+\nu}\left(\delta w_r (r^2R(r^2R-4) +2\kappa (w^2-1)^2) \right)\\
&+4r^2 R^2 e^{\lambda+\nu} \delta w (3w^2-1)+
16\kappa rR e^{\lambda+\nu} \delta w w_r\, w(w^2-1) =0\, ;\\[4pt]
&4e^\lambda r^3R \delta v_{tt} + r^2 R e^{\lambda+\nu} w \delta u(r^2R-4) +
4r R e^{\lambda+\nu} \delta v (w^2-1) )\\
&+e^{\lambda+\nu}\delta v_r \left(r^2R(r^2R-4) +2\kappa(w^2-1)^2 \right)
+2 \kappa e^{\lambda+\nu} \delta u \, w(w^2-1)^2 \\
&+4 e^\nu r^2R\left(\delta v_r - 2r w_r \delta u - r w \delta u_r - r \delta v_{rr} +w \delta u
\right) = 0\, ;\\[4pt]
&2r^2\delta u_{tt} + 4e^\nu w(\delta v_r + w \delta u) -4 e^\nu w_r \delta v = 0
\end{split}
\ee
Note that the equation for the fluctuations  $\delta w$  is decoupled from the
other two equations, which involve $\delta v$ and $\delta u$. The first equation defines the even
parity fluctuations, while the other two equations correspond to the odd parity fluctuations.

Now we can suppose the fluctuations are harmonic, i.e.
\be
\begin{split}
\delta \nu(r,t) &= \delta \nu (r) e^{i\omega t}\, , \quad
\delta \lambda(r,t) = \delta \lambda (r) e^{i\omega t}\, ; \\
\delta w(r,t) &= \delta w(r) e^{i\omega t}\, , \quad
\delta u(r,t) = \delta u(r) e^{i\omega t}\, , \quad
\delta v(r,t) = \delta v(r) e^{i\omega t}\, .
\end{split}
\ee
Thus, the eigenvalue equation for the even-parity perturbations is
\be
A \delta_w{rr} +B \delta w_r + C \delta w = \omega^2 \delta w \, ,
\ee
and the odd-parity perturbations are described by the equations
\be
\begin{split}
D \delta_w{rr} &+ E \delta v_r + F \delta v + G \delta u_r + H \delta u = \omega^2 \delta v \, ,\\
I \delta v_r &+J \delta v + K \delta u = \omega^2 \delta u
\label{eigenvalues}
\end{split}
\ee
where $A\dots K$ are the following functions of unperturbed solution:
\be
\begin{split}
A&=D=-e^{\nu-\lambda}\, , \quad B=E = \frac{e^\nu}{4r^3}\left(r^4 R + 4r^2(e^{-\lambda} -1 )
+\frac{2\kappa}{R} + \frac{2\kappa w^2(w^2-2 ) }{R}\right)\, ,\\
C&=\frac{\kappa e^\nu}{2r^4R^2} \biggl(
\frac{2r^2 R^2 (3w^2\! -\!1) }{\kappa}\! +\! w_r^2 \left[
2\kappa\! +\! r^2R(r^2R\! -\!4)\! +\!2\kappa w^2 (w^2\!-\!2) \right]
\!+\! 8rR \, w w_r (w^2\!-\!1)
\biggr)\, ,\\
F&=\frac{e^\nu (w^2-1)}{r^2}\, , \quad G= - e^{\nu-\lambda} w \, ,\\
H&=\frac{e^\nu }{4r^3 R}\left(w [2\kappa + r^2 R (r^2 R -4) +2\kappa w^2 (w^2-2)  ]
+ 4 r^2 R e^{-\lambda} (w - 2r w_r) \right)\, , \\
I&=\frac{2e^\nu w}{r^2}\, , \quad J = -\frac{2e^\nu w_r}{r^2}
\, , \quad K = \frac{2e^\nu w^2}{r^2} \, .
\end{split}
\ee

The eigenvalue problem \re{eigenvalues} can be solved numerically, the case of imaginary eigenvalues $\omega^2< 0$
corresponds to the exponential growth of the perturbation, i.e. instability of the original solution.
We found that for the odd-parity perturbations
the results are similar with the corresponding situation in the EYM model \cite{Bjoraker:1999yd,Bjoraker:2000qd},
the number of unstable modes is equal to the number of nodes
of the profile function $w(r)$. Thus, the nodeless solutions are stable with respect to
the odd-parity perturbations.

The situation is much simpler for the even-parity perturbations, it turns out, that for any values
of the gravitational coupling constant $\kappa$ and the shooting parameter $b$, the eigenvalues of the
problem \re{eigenvalues} are real, i.e. the solutions are stable with respect to the
even-parity perturbations. Fig.~\ref{f-3} shows for example, the evolution of ten lowest eigenvalues of the
even-parity perturbations as the shooting parameter $b$ varies, all eigenvalues remain real.

\begin{figure}[hbt]
\lbfig{f-3}
\begin{center}
\includegraphics[width=.6\textwidth, angle =0,trim = 60 20 80 50, clip = true]{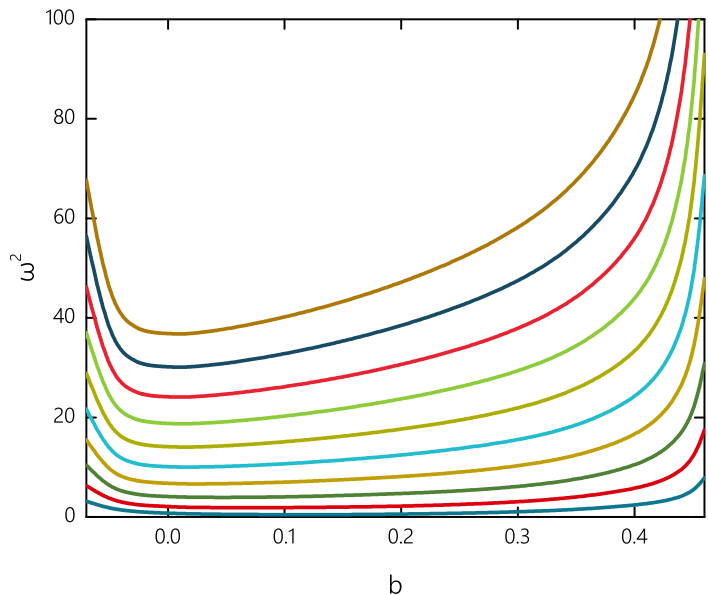}
\end{center}
\caption{\small The eigenvalues of ten lowest even-parity perturbations of the
$R^2$ EYM solutions are shown as functions
of the parameter $b$ for $\kappa=1$ and $R=-1$.
}
\end{figure}

\section{Conclusions}
The objective of this work was to investigate properties of new regular
solutions of the $\mbox{SU}(2)$ Yang-Mills
theory, coupled to the pure $R^2$ gravity. We found a family of non-trivial spherically symmetrical
solutions with non-vanishing magnetic charge, which
generalize the usual Bartnik-McKinnon solitons in the EYM theory.
We found that the $R^2$ EYM model has only solutions with no node in the gauge field profile function,
or with a single node, there is no solutions with multiple nodes.
Similar to the BM solutions in asymptotically AdS$_4$ spacetime, the nodeless solutions
are stable with respect to linear perturbations.

We remark that the scale invariant $R^2$ EYM model is very different from the generalizations of gravity, which also
include the usual linear in curvature term. The $R+R^2$ model is also ghost-free, however it is equivalent to the
conventional Einstein gravity with an additional scalar field. Further, we found that such a model supports only
regular solutions, for which the curvature scalar is zero and $T_{\mu\nu}=0$.

The scale invariance of the $R^2$ model with the non-Abelian matter fields will be broken when the $R^2$ gravity
will be coupled to the Yang-Mills-Higgs system with symmetry breaking potential. We expect the properties of the
corresponding monopole solutions will be different from the standard gravitating monopoles in asymptotically
AdS$_4$ spacetime. Another  direction for further work is to investigate $R^2$ EYM non-Abelian configurations with both
electric and magnetic charges.

\section*{Acknowledgements}
We are grateful to Piotr Bizon,  Burkhard Kleihaus, Jutta Kunz and Michael Volkov for inspiring and valuable discussions.
Y.S. gratefully acknowledges support from the Ministry of Education and Science
of Russian Federation, project No 3.1386.2017, JINR Heisenberg-Landau Program of collaboration Oldenburg-Dubna,
and DFG (Grant LE 838/12-2). We would like to thank the Department of Physics, Carl von
Ossietzky University of Oldenburg, for its kind hospitality.

\end{document}